\documentclass[12pt,reqno]{article}
\usepackage{amsmath,amsfonts,amssymb}

\usepackage{amsfonts}
\usepackage{amssymb}
\def\a{\alpha} 
\def\b{\beta}

\def\d{\delta}

\numberwithin{equation}{section}

\title{A Unified Invariant Formulation, by Frames, from General Relativity to the Atomic Scale}

\author{Shmuel Kaniel\\
{Institute of Mathematics, Hebrew University of
   Jerusalem }\\
{kaniel@math.huji.ac.il}
}
 \begin{document}
\maketitle
\begin{abstract}
The aim of this article is to advocate the formulation of the basic laws of Physics by frames $\Phi$, i.e. quadruples of exterior differential one-forms. These are invariant w.r. to  any diffeomorphism. The hyperbolic $*$ is modified to $* _\Phi$ which is invariant. The basic operator is a modification of the Hodge-de Rham Laplacian $\square=d*d*+*d*d$ to $  \square_\Phi=d*_\Phi d*_\Phi+*_\Phi d*_\Phi d$. 
The basic equation is motivated by the Einstein equation in nonempty space.  Einstein's $G_{\mu\nu}$ is substituted by $\square_\Phi$. 
The  field equation is $\square_\Phi\Phi=\lambda(x)\Phi$, where $\lambda(x)$ is a function of  the entries of $\Phi$ and their first order derivatives.  Kaniel and Itin \cite{4} showed that a similar equation  results in a complete alternative to the field equation of General Relativity in vacuum. Then first order linear approximation  of $\square_\Phi$ is considered. This way, natural invariant formulation of Maxwell equations is exhibited. After that  invariant formulation of Schroedinger equation (classical and relativistic) and Dirac equation is derived. 
The frame-field equation yields a derivation of Newtonian (Einstein) law of attraction without recourse to the geodesic postulate. Coulomb law is also derived. 
\end{abstract}
\section{Introduction}
The aim of this article is to advocate the formulation of the basic laws of physics by frames $\Phi$, i.e., quadruples of  exterior differential one-forms, invariant entities. One invariant field equation that ranges from the Universal to the Atomic Scale is exhibited. Different laws are characterized by  assumptions on the energy content in space. The article is motivated by General Relativity,  in particular by the Einstein equation in nonempty space. His tensor $G_{\mu\nu}=R_{\mu\nu}-\frac 12 g_{\mu\nu}R$  is replaced by $\square_\Phi$ to be defined in the sequel. 

A. Einstein, in the general theory of relativity \cite{1} postulated that
\begin{itemize}
\item[1.] The world is a four dimensional manifold
\item[2.] Gravitation is a construct of a Riemannian manifold.
\item[3.] The field equation, in Vacuum, is $R^\b_\a=0$, where  $R^\b_\a$ is Ricci's tensor. 
\item[4.] The geodesic postulate. Pointlike massive bodies move on geodesics of the metric.
\item[5.] The form of the equations should be independent of the coordinate system. 
\end{itemize}
Once it is postulated that the world of gravity is Riemannian then, in principle, the only plausible choice of an invariant construct for a field equation is Ricci's tensor or a modification of it. 
Consequently, any attempt to define a novel invariant field equation should be based on a different construct.
In this article the construct is taken  to be Cartan's frame \cite{3}, \cite{5}, a quadruple of four differential one-forms
\begin{equation}\label{1.0}
\Phi=\Phi^\a=\Phi^\a_\b dx^\b
\end{equation}
Notation: A Greek letter index ranges over $(0,1,2,3)$. A Roman letter index rangers over $1,2,3$. A repeated index is subject  to Einstein summation convention. Derivatives are denoted by bar-index:$\frac{\partial f}{\partial x^\a}=f_{|\a}$.

A frame $\Phi$ yields the metric $g$ by 
  \begin{equation}\label{1.1}
g_{\mu\nu}=\eta_{ab}\Phi^a_\mu\Phi^b_\nu \end{equation}
  where $\eta_{ab}=diag(-1,1,1,1)$ the Lorentzian metric tensor. 
The frame is assumed to be complex. Mass and forces, including the electromagnetic forces are taken to be real (cf \cite{L-L}). 
For a complex frame, 
 \begin{equation}\label{1.1x}
g_{\mu\nu}=Re\left(\eta_{ab}\Phi^a_\mu\Phi^b_\nu \right)\,.
\end{equation}
Recall the operator $d$ acting on exterior forms
 \begin{equation}\label{1.2}
  d\left(f(x)dx^{\a_1}\wedge\cdots\wedge dx^{\a_k}\right)=f(x)_{|\a} dx^\a\wedge dx^{\a_1}\wedge\cdots\wedge dx^{\a_k}
  \end{equation}
   \begin{equation}\label{1.3}
  d\left(f(x)dx^{\a_1}\wedge dx^{\a_2}\right)=-d\left(f(x)dx^{\a_2}\wedge dx^{\a_1}\right)
  \end{equation}
  consequently for any form $W$
   \begin{equation}\label{1.4}
  d^2W=0
  \end{equation}
  A frame $\Phi$ is defined by its structural equations \cite{3}.
\begin{equation}\label{1.5}
  d\Phi^\a=\chi^\a_{\lambda\mu}\Phi^\lambda\wedge\Phi^\mu
  \end{equation}
together with
\begin{equation}\label{1.6}
dH=H_{|\a}\Psi^\a_\b\Phi^\b
 \end{equation}
 where $H$ is an arbitrary function. $\Psi^\a_\b$ is an inverse of $\Phi^\a_\b$. 
 For $\Phi$ defined by (\ref{1.0})
 \begin{equation}\label{1.7}
d\Phi^\a=\Phi^\a_{\b|\gamma}
\Psi^\gamma_\lambda\Psi^\b_\mu\Phi^\lambda\wedge\Phi^\mu=\chi^\a_{\lambda\mu}\Phi^\lambda\wedge\Phi^\mu
 \end{equation}
 The structural equations determine the frame. The $\chi^\a_{\lambda\mu}$ are scalars, invariant under any diffeomorphism.
 Each $\Phi^\a$ represents a material distribution in space. 
 
 In the sequel, the basic field equations will be derived, a Lagrangian will be exhibited, the fields of point particles will be constructed. The fields of mass and electric charge will be identified. Next, the linearized equations will be considered.  Then , invariant Maxwell equations are formulated using the invariant operators $d$ and$*_\Phi$, to be defined in the sequel. 
Schroedinger and Dirac equations, being linear, will be reformulated by the linearized frame field equations. Subatomic physics is not considered in this paper. 
Newton (Einstein) law of attraction will be deduced by the field equations without recourse to the geodesic postulate. A number of simplifying  assumptions are made in the derivation of Newton and Coulomb laws.    All the equations evaluated in this article are invariant w.r. to any diffeomorphism.  The lhs of the field equation, defined on a four dimensional manifold, is the same for all bodies. The rhs depends on the energy content at any point. 

It is suggested to extend the derivation by frames on a four dimensional space to laws of electrodynamics and  quantum mechanics. Elementary particles may also be considered, by taking the space of frames to be a representation space of the Lorentz group. As such, this space contains a wealth of irreducible representations that may describe elementary particles. 

\section{The basic field equation}
Recall  the hyperbolic star operator -- $*$. Denote $dx^0=cdt$. Consequently
    \begin{equation}\label{2.1}
  *\left(dx^{\a_1}\wedge\cdots\wedge dx^{\a_k}\right)=(-1)^ldx^{\b_1}\wedge\cdots\wedge dx^{\b_{n-k}}
  \end{equation}
  where $\a_1,\cdots,\a_k,\b_1,\cdots,\b_{n-k}$ is an even permutation of $(0,1,2,3)$. $l=1$ is zero if one of the $\b_i$, $l=0$ otherwise. 

The Hodge-de Rham Laplacian is defined by
  \begin{equation}\label{2.2}
  \square=d*d*+*d*d
   \end{equation}
   On functions and 1-forms 
    \begin{equation}\label{2.3}
    \square f=\frac{\partial^2 f}{\partial x_0^2}-\triangle f;\qquad 
     \square (fdx^\a)=\square f dx^\a
       \end{equation}
       The principal definition in this article is that of $*_\Phi$. The coefficients of the forms in this article are denoted by $exp(-f)$ or $exp(g)$ etc. $f$ and $g$ are taken to be small. 
The derivatives of $f$ and $g$ will be referred to as first order. 
The  second order derivatives, say $exp(-f)_{|1|1}$ is composed of two terms $-f_{|1|1}exp(-f)$, which is first order and $-f_{|1}f_{|1}exp(-f)$ which is second order. $*_\Phi$ will be defined as follows: For $f$, a first order,   
          \begin{equation}\label{2.5}
  *_\Phi\left(f\Phi^{\a_1}\wedge\cdots\wedge \Phi^{\a_k}\right)=(-1)^lf\Phi^{\b_1}\wedge\cdots\wedge \Phi^{\b_{n-k}}
  \end{equation}
 where $\a_1,\cdots,\a_k,\b_1,\cdots,\b_{n-k}$ is an even permutation of $(0,1,2,3)$. $l=1$ if zero is one of the $\b_i$, $l=0$ otherwise.  If $f$ is second order then $l=0$.

$*_\Phi$ and, consequently, $\square_\Phi$ are invariant. There is  freedom to modify each equation of (\ref{2.5}), separately to 
 \begin{equation}\label{2.5x}
  *_\Phi\left(f\Phi^{\a_1}\wedge\cdots\wedge \Phi^{\a_k}\right)=\mu(\a_1,\cdots, \a_k)(-1)^lf\Phi^{\b_1}\wedge\cdots\wedge \Phi^{\b_{n-k}}
  \end{equation}
  retaining the invariance of $*_\Phi$ and $\square_\Phi$.In this article we'll define 
 \begin{equation}\label{2.6}
      *_\Phi \Phi^0=-\frac 13 \Phi^1\wedge\Phi^2\wedge\Phi^3
        \end{equation}
        Due to (\ref{2.5x}), there is  great freedom in the definition of  $*_\Phi$.  
 (\ref{2.6}) is motivated by the desire to make $\square_\Phi$ as close to $\square$ as possible. If, for some reason, the identity $*_\Phi*_\Phi=-1$ is desired then the equation 
\begin{equation}\label{2.7}
     *_\Phi\Phi^1\wedge\Phi^2\wedge\Phi^3=-3\Phi^0
        \end{equation}
        will do it. (\ref{2.7}) is never used in this paper. 
        
       Define
       \begin{equation}\label{2.4}
       \square_\Phi=d*_\Phi d*_\Phi+*_\Phi  d*_\Phi d
        \end{equation}
   
The basic field equation is 
\begin{equation}\label{2.9}
      \square_\Phi\Phi=\lambda(x)\Phi
        \end{equation}
where $\lambda(x)$ is a source term, composed of functions and first order derivatives. 
The  frame $\Phi$ takes place of the metric $g_{\mu\nu}$. $\square_\Phi$ substitutes Einsteins $G_{\mu\nu}=R_{\mu\nu}-\frac 12 g_{\mu\nu}R$. 
The frame $\Phi$ is composed of two sub-frames, $\Phi^0$ and $\Phi^j$. $\square_\Phi\Phi^0$ and $\square_\Phi\Phi^j$ consist of two  invariant forms.  (\ref{2.6}) is  needed only for $\square_\Phi\Phi^0$. In most cases the two resulting equations are duplicates.

In \cite{4}, it is proven that equation (\ref{2.4}) is the Euler equation of appropriate Lagrangian. The $*$ in \cite{4} is $*_\Phi$ in this paper. Define $\delta=*_\Phi d*_\Phi$. The Lagrangian will be 
\begin{equation}\label{2.10}
      L=\eta_{\a\b}\left(d\Phi^\a\wedge d\Phi^\b+\delta \Phi^\a\wedge \delta\Phi^\b\right)\,.
        \end{equation}
In order to get the field  equations of \cite{4}, one takes variations of the 1-forms that commute with  $*_\Phi$. It follows that the volume element $*_\Phi=\Phi^0\wedge\Phi^1\wedge\Phi^2\wedge\Phi^3$ is preserved. 

 \section{The Linearized Equation}
 $\square_\Phi\Phi$ is  a very complicated object, cf \cite{5}. Consequently, let us compute it's linearization $L\square_\Phi\Phi$. It is invariant to the first order. Recall that A. Einstein \cite{1} had computed, first, the linearized equation. 
  For that it is assumed that $\Phi^\a_\b=\d^\a_\b+f^\a_\b$, where $f^\a_\b$ is assumed to be small. Consequently, in the course of computation, products of $f^\a_\b$ and their derivatives are omitted. After a $d$ operation (which is linear), the action of the operator $L*_\Phi$ on $d\Phi$ is equivalent to the action of the linear $*$, i.e., $*d\Phi$.   
Thus  
 \begin{equation}\label{7.1}
(L*_\Phi) Ld  (L*_\Phi)  d=*d*d\,.
 \end{equation}
 The equation 
 \begin{equation}\label{7.4}
L\square_\Phi\Phi=\square\Phi\,
  \end{equation}
 holds provided that  
the equation 
 \begin{equation}\label{7.2}
d (L*_\Phi) d  (L*_\Phi)\Phi=d*d(L*_\Phi)\Phi=d*d*\Phi\,
  \end{equation} 
  holds. 
This is verified provided that the equation
  \begin{equation}\label{7.3}
d  (L*_\Phi)\Phi=d*\Phi\,.
  \end{equation}
  is verified.

 Let us compute, first, the linearization of  equation (\ref{2.9}) for a stationary and diagonal frame. The equation will be 
  \begin{equation}\label{7.5}
 L\square_\Phi\Phi=0\,.
  \end{equation}
 Denote $x^0=ct$
  \begin{equation}\label{7.5a}
 \Phi^0=[1-f(r)]dx^0\,,\qquad \Phi^j=[1+g(r)]dx^j
  \end{equation}
 To the first order, 
\begin{equation}\label{7.5x}
\Phi^1\wedge\Phi^2\wedge\Phi^3=(1+3g)dx^1\wedge dx^2\wedge dx^3 \end{equation}
 \begin{equation}\label{7.5xx}
*\Phi^0=(1-f)dx^1\wedge dx^2\wedge dx^3\end{equation}
 and
 \begin{equation}\label{7.5xxx}
d*_\Phi\Phi^0=d*\Phi^0=0
\end{equation}
 Consequently, by (\ref{7.1}), if $\square\Phi^0=0$ then $L\square_\Phi\Phi^0=0$. 
 
 By Appendix A, $f=g$. For that, the linearized equations is not enough. The exact solution is needed. 
 
With (\ref{2.6}), 
 \begin{equation}\label{7.6x}
d*_\Phi\Phi^0=d\left(-g dx^1\wedge dx^2\wedge dx^3 \right)=d(-fdx^1\wedge dx^2\wedge dx^3 )=d*\Phi^0
\end{equation}
So again 
 \begin{equation}\label{7.6xx}
L\square_\Phi\Phi^0=\square\Phi^0 \end{equation}
  Thus, 
   \begin{equation}\label{7.7}
\Phi^0=(1-\frac mr)dx^0
  \end{equation}
  Now
  \begin{eqnarray}\label{7.7x}
*_\Phi\Phi^1&=&\Phi^0\wedge\Phi^2\wedge\Phi^3=(1+2g-f)dx^0\wedge dx^2\wedge dx^3\nonumber\\
  *\Phi^1&=&(1+g)dx^0\wedge dx^2\wedge dx^3
  \end{eqnarray}
  By Appendix A,  $g=f$.  Thus $*_\Phi\Phi^1=*\Phi^1$ and $\square_\Phi\Phi^1=\square\Phi^1.$
  Thus
    \begin{equation}\label{7.8}
\Phi^j=(1+\frac mr)dx^j
  \end{equation}
  The line element will be 
   \begin{equation}\label{7.9}
ds^2=-(1-\frac {2m}r){dx^0}^2+(1+\frac {2m}r)dr^2\end{equation}
  The linearized Einstein line element.
  
  The essential equation $f=g$ is implied by  the definition of $\square_\Phi\Phi$. The operator $\square$ by itself is not enough.

\section{Schroedinger equation, relativistic and non-relativistic and Dirac equation}
In this section we will show that the equations above can be reformulated in terms of frames. 
The equations are linear. Thus, the linearized form of (\ref{2.9}) will be assumed to be 
  \begin{equation}\label{8.1}
L\square_\Phi\Phi^\a= L\square_\Phi(\Phi^\a-dx^\a)=g(x)(\Phi^\a-dx^\a)
\end{equation}
 (\ref{8.1}) is invariant to the first order. 
By (\ref{7.5a}) and (\ref{7.10})  
   \begin{equation}\label{8.3}
L\square_\Phi\Phi=\square\Phi
\end{equation}
$f$ in (\ref{7.5a}) and $\phi$ in (\ref{7.10}) depend on the rhs of (\ref{8.1}) which in turn depends on the particular equation. $g(x)$ represents the matter content of the mass and charges pertaining to a particle satisfying (\ref{8.1}).

 It is enough to show that Schroedinger and Dirac equations are equivalent to (\ref{8.1}). 
 
Also here it will be seen that $L\square_\Phi\Phi^\a$ is composed of a duplicate:  $L\square_\Phi\Phi^0$ and $L\square_\Phi\Phi^j$. 

The following formulae are taken from \cite{2}. The non relativistic Schroedinger equation by (16.6) of \cite{2} with minor rearrangement is 
  \begin{equation}\label{8.9}
-\frac 1{r^2}\frac d{dr}\left(r^2\frac{dR}{dr}\right)+\frac{l(l+1)R}{r^2}=\frac{2\mu}{\hbar^2}\left(E+\frac{Ze^2}r\right)R
\end{equation}
  $\mu$ is the mass and $e$ is the charge of the electron. The solution of (\ref{8.9}) holds for discrete values of $E$
  \begin{equation}\label{8.10}
-|E_n|=-\mu\frac {Z^2e^4}{2\hbar^2n^2}
\end{equation}
  By (51-14) of \cite{2} the relativistic Schroedinger equation is
  
   \begin{eqnarray}\label{8.11}
&&-\frac 1{r^2}\frac{d}{dr}\left(r^2\frac{dR}{dr}\right)+\frac{l(l+1)}{r^2}R=\frac{(E-e\phi)^2-m^2c^4}{\hbar^2c^2}R\nonumber\\
&=&(E^2-2EZe^2\frac 1r+Z^2e^4\frac 1{r^2}-m^2c^4)R
\end{eqnarray} 
where $e\phi=-\frac{Ze^2}{r}$. By(51.15) of \cite{2} the non-dimensional form of (\ref{8.11}) is
 \begin{equation}\label{8.12}
\frac 1{\rho^2}\frac{d}{d\rho}(\rho^2\frac{dR}{d\rho})+\left[\frac\lambda\rho -\frac 14-\frac{l(l+1)-\gamma^2}{\rho^2}\right]R=0
\end{equation}
where
\begin{eqnarray} \label{8.13}
&&\rho=\a r, \qquad \gamma=\frac{Ze^2}{\hbar c}\nonumber\\
&&\a^2=\frac{4(m^2c^4-E^2)}{\hbar^2 c^2},\qquad \lambda=\frac{2E\gamma}{\hbar c\a}
\end{eqnarray} 
A solution of (\ref{8.11}) holds only if 
 \begin{equation}\label{8.14}
E=mc^2[1-\frac{\gamma^2}{2n^2}-\frac{\gamma^4}{2n^4}(\frac n{l+\frac 12}-\frac 34)]
\end{equation}
The lhs of (\ref{8.9}) and (\ref{8.11}) is $\square R$ where $R(x)=T(r)Y^{l,j}(\theta,\varphi)$. 

$Y^{l,j}(\theta,\varphi)$ are the spherical harmonics. Thus, $g(x)$ in (\ref{8.1}) is the rhs of (\ref{8.9}) and (\ref{8.11}), respectively.

Now let us show that Dirac equation, too, can be expressed by (\ref{8.1}). Start with equation (53.15) of \cite{2}
 \begin{equation}\label{8.15}
(E-mc^2-V)F+\hbar c \frac{dG}{dr}+\frac{\hbar c k}r G=0
\end{equation}
\begin{equation}\label{8.16}
(E+mc^2-V)G-\hbar c \frac{dF}{dr}+\frac{\hbar c k}r F=0
\end{equation}
Differentiate and multiply by $(\hbar c)^{-1}$
 \begin{equation}\label{8.17}
\frac{d^2G}{dr^2}=-(\hbar c)^{-1}(E-mc^2-V)\frac{dF}{dr}+(\hbar c)^{-1} \frac{dV}{dr}F-\frac kr\frac{dG}{dr}+\frac k{r^2}G
\end{equation}
 \begin{equation}\label{8.18}
\frac{d^2F}{dr^2}=(\hbar c)^{-1}(E+mc^2-V)\frac{dG}{dr}-(\hbar c)^{-1} \frac{dV}{dr}G+\frac kr\frac{dF}{dr}-\frac k{r^2}F
\end{equation}
Substitute for $\frac{dF}{dr}$ and $\frac{dG}{dr}$ in (\ref{8.17}) and (\ref{8.18}) the values from (\ref{8.15}) and (\ref{8.16}). 
\begin{equation}\label{8.19}
\frac{d^2G}{dr^2}=(\hbar c)^{-2}(m^2c^4-E^2+2EV-V^2)G+(\hbar c)^{-1} \frac{dV}{dr}F+\frac {k(k+1)}{r^2}G
\end{equation}
 \begin{equation}\label{8.20}
\frac{d^2F}{dr^2}=(\hbar c)^{-2}(m^2c^4-E^2+2EV-V^2)F-(\hbar c)^{-1} \frac{dV}{dr}G+\frac {k(k-1)}{r^2}F
\end{equation}
Express $$V=e\phi=-\frac{Ze^2}r=-(\hbar c)\gamma \frac 1r$$
write (\ref{8.17}) and (\ref{8.18}) in nondimensional form using  (\ref{8.13})
\begin{equation}\label{8.21}
-\frac{d^2G}{d\rho^2}=\left[\frac \lambda \rho-\frac 14 -\frac {k(k+1)-\gamma^2}{\rho^2}\right]G+\frac{\gamma F}{\rho^2}
\end{equation}
 \begin{equation}\label{8.22}
-\frac{d^2F}{d\rho^2}=\left[\frac \lambda \rho-\frac 14 -\frac {k(k-1)-\gamma^2}{\rho^2}\right]F-\frac{\gamma G}{\rho^2}
\end{equation}
Let us compute coefficient $s$ so that for $H=aG+bF$ 
\begin{equation}\label{8.23}
-\frac{d^2H}{d\rho^2}=\left[\frac \lambda \rho-\frac 14 -\frac {k^2-\gamma^2+s}{\rho^2}\right]H
\end{equation}
$a$, $b$ and $s$ should satisfy 
\begin{equation}\label{8.24}
a(-kG+\gamma F)+b(kF-\gamma G)=s(aG+bF)
\end{equation}
$s$ is an eigenvalue of the matrix 
$$T=\left( \begin{array}{cc}
-k & -\gamma  \\
\gamma  & k  \end{array} \right)$$
Thus $s^2=k^2-\gamma^2$, take the positive root. Set $H=\rho^2 K$. By (\ref{8.23}) 
 \begin{equation}\label{8.25}
\triangle K=-\frac 1{\rho^2}\frac{d^2(\rho^2K)}{d\rho^2}=\left[\frac \lambda \rho-\frac 14 -\frac {k^2-\gamma^2+(k^2-\gamma^2)^{1/2}}{\rho^2}\right]K
\end{equation}
 as needed in (\ref{8.1}). 

Eq. (\ref{8.25}) is the result of  mathematical manipulation of Dirac equation. 
Let us show that the computations of the energy levels by (\ref{8.25}) agrees with Dirac's 
\begin{equation}\label{8.26}
E=mc^2\left[1-\frac{\gamma^2}{2n^2}-\frac{\gamma^4}{2n^4}\left(\frac n{|k|}-\frac 34\right)\right]
\end{equation}
Indeed, as in Schroedinger relativistic equation, up the fourth order in $\gamma$, it follows that 
\begin{equation}\label{8.27}
\lambda=n'+s+1
\end{equation}
\begin{equation}\label{8.28}
E=mc^2\left(1+\frac {\gamma^2}{\lambda^2}\right))^{-1/2}=mc^2\left(1-\frac 12 \frac {\gamma^2}{\lambda^2}+\frac 38 \frac {\gamma^4}{\lambda^4}\right)\,.
\end{equation}
Here
\begin{equation}\label{8.29}
s(s+1)=k^2-\gamma^2+(k^2-\gamma^2)^{1/2}\,,\qquad s=(k^2-\gamma^2)^{1/2}
\end{equation}
It is enough to compute $1/\lambda^2$ to the second order in $\gamma$. 
\begin{equation}\label{8.30}
\frac 1{\lambda^2}=\frac 1{n^2}\left(1+\frac{\gamma^2}{nk}\right)\,, \qquad n=n'+k+1\,.
\end{equation}
The substitution of (\ref{8.30}) in (\ref{8.28}) results in (\ref{8.26}).

Let us show that we can get the energy levels of Dirac equation also by modifying the energy term of  Schroedinger relativistic equation i.e. adding energy due to the spin of the electron. Let us modify $\gamma$ to $\hat{\gamma}$. 
\begin{equation}\label{8.31}
E=mc^2\left[1-\frac{\hat{\gamma}^2}{2n^2}-\frac{\hat{\gamma}^4}{2n^4}\left(\frac n{l+\frac 12}-\frac 34\right)\right]
\end{equation}
This has to be agree with (\ref{8.30}) for $l=k$. Let $\hat{\gamma}^2={\gamma}^2(1+{\gamma}^2\a^2)$. Compare (\ref{8.31}) to (\ref{8.26}) up to the fourth order $ \gamma$.
\begin{equation}\label{8.32}
\a^2+\frac 1{n^2}\left(\frac n{l+\frac 12}-\frac 34\right)=\frac 1{n^2}\left(\frac n{l}-\frac 34\right)
\end{equation}
Thus 
\begin{equation}\label{8.33}
\a^2=\frac 1{2n}\cdot \frac 1{l\left(l+\frac 12\right)}
\end{equation}
Eq. (\ref{8.1}) is satisfied by taking 
for the frame (\ref{7.5a})  $f=R$ and, for the frame  (\ref{7.10}) $\phi=R$. This is substituted in (\ref{8.9}), (\ref{8.11}) and (\ref{8.12}), respectively.  Likewise  $f=K$, or $\phi=K$, respectively is substituted  in (\ref{8.25}). 
\section{ The frame of a stationary, spherically symmetric field} 
Let $\Phi$ be a complex frame. Wlog one may take the frame to be diagonal so that 
\begin{eqnarray}\label{3.1}
\Phi^0&=&exp(-f)dx^0\\ \label{3.2}
\Phi^j&=&exp(g)dx^j
\end{eqnarray}
It is assumed that $x^0=ct$.

$\square_\Phi\Phi^0$  and $\square_\Phi\Phi^j$ are derived in  Appendix A. There it is shown that 
$f=g$ is needed for equation (\ref{2.9}) to hold. Thus
\begin{eqnarray}\label{3.4}
\Phi^0&=&exp(-f)dx^0\\ \label{3.5}
\Phi^j&=&exp(f)dx^j
\end{eqnarray}
By (\ref{A.11}) and (\ref{A.19}) 
\begin{eqnarray}\label{3.6}
\square_\Phi\Phi^0&=&(-f_{|i|i}+f_{|i}f_{|i})exp(-2f)\Phi^0\\ \label{3.7}
\square_\Phi\Phi^j&=&(-f_{|i|i}+f_{|i}f_{|i})exp(-2f)\Phi^j
\end{eqnarray}
 (\ref{3.6}) and   (\ref{3.7}) imply separately  (\ref{2.9}). 

By (\ref{2.9}) it follows that $f_{|i|i}=0$.  Thus 
\begin{equation}\label{3.8}
f=\frac{m+iq}r
\end{equation} 
\begin{equation}\label{3.9}
\Phi^0=exp\left(-\frac{m+iq}r\right)dx^0
\end{equation} 
\begin{equation}\label{3.10}
\Phi^j=exp\left(\frac{m+iq}r\right)dx^j
\end{equation} 

\begin{equation}\label{3.10x}
\square_\Phi\Phi^0=-f_{|i}f_{|i}\Phi^0, \qquad \square_\Phi\Phi^k=-f_{|i}f_{|i}\Phi^k
\end{equation} 
This is a duplication.  
For a real frame, a similar equation  is dealt with by Kaniel and Itin in \cite{4}, see also \cite{8}. 
 There, a closed solution  is computed. 
 
 By (\ref{1.1}) the frame (\ref{3.9}) or (\ref{3.10}) yields, for $q=0$,  the Rosen metric \cite{7}
 \begin{equation}\label{11}
ds^2=e^{-2m/r}dt^2+e^{2m/r}\left(dx^2+dy^2+dz^2\right)\,.
 \end{equation}
 The solution (\ref{11}) is essentially different from the Schawarzschild solution. It's curvature is $$\frac{2m^2}{r^4}e^{-2m/r}\ne 0\,.$$
 The singularity is a point singularity unlike Schawarzschild radius. Nevertheless, black holes do exist. The two metrics are indistinguishable with respect to three classical experimental tests. Both theories  rely on the geodesic hypothesis. Thus the three metrics, Schawarzschild, Kaniel and Itin and Kaniel (this article) share the second order terms of $\Phi^0$ and the first order terms of $\Phi^j$. These are  the only terms that count toward the verification of the experimental tests. Recall that A. Einstein had computed, first, the linearized equation \cite{1}. 
 \section{Interaction of two bodies. Newton and Coulomb laws}
 Let two bodies (particles) move on the trajectories $\a^{(l)}(x^0)\,, l=1,2$.  Choose the center so that $\a^{(2)}(x^0)=0$. Suppose that for a time spot $y^0$ the first  body is, momentarily, at rest i.e. $\dot{\a} ^{(1)}(y^0)=0$ ($\dot{\a} ^{(l)}=\frac{d{\a} ^{(l)}}{dx^0}$.) 
 Take the frame of each particle to be defined by (\ref{3.4}) and (\ref{3.5}).  $f^{(l)}$ are defined by (\ref{3.8}) where $r=r^{(l)}=\{(x^j-\a^{(l,j)})^2\}^{1/2}$. Consequently take the frame pertaining to each particle to be defined by (\ref{3.9}) and (\ref{3.10}) where, again, $r=r^{(l)}$. 

\underline{Ansatz 1} The combined  frame of the two particles is taken to be ,  approximately,   the product of the frames (\ref{3.4}) and (\ref{3.5}), thus 
 \begin{equation}\label{4.1}
f=f^{(1)}+f^{(2)}\,,
 \end{equation}
 where $f^{(1)}$ and $f^{(2)}$ solve (\ref{3.6}) and (\ref{3.7}) leading to (\ref{3.9}) and (\ref{3.10}). 
 \begin{equation}\label{4.2}
\Phi^0=exp\{-(f^{(1)}+f^{(2)})\}dx^0=exp(-f)dx^0\,.
 \end{equation}
 \begin{equation}\label{4.3}
\Phi^j=exp\{f^{(1)}+f^{(2)}\}dx^j=exp(f)dx^j\,.
 \end{equation}
 Let $f^{(i)}$ generate $\Phi^{(i)}$ respectively. 
 
 \underline{Ansatz 2} 
 \begin{equation}\label{4.3x}
\square_\Phi\Phi=\square_\Phi\Phi^{(1)}+\square_\Phi{\Phi}^{(2)}\,.
 \end{equation}
  It is assumed that the field of each particle is not affected by the existence of the other particle. 
  
  The computation of $\square_\Phi\Phi$ for the time dependent case is preformed in Appendix B. Evaluate (\ref{B.13}) and (\ref{B.20}) for $f^{(1)}, f^{(2)}$ and $f=f^{(1)}+f^{(2)}$. Recall that $\a^{(2)}= 0$, $\a^{(1)}(y^0)= 0$. 
Let us, further, approximate (\ref{3.5}), (\ref{3.6}),  
(\ref{B.13}) and (\ref{B.20}) by substituting 1 for the exponents. 
Eq. (\ref{4.3x}) will turn out to be 
 \begin{eqnarray}\label{4.6y}
&&-f^{(1)}_{|0|0}+(f^{(1)}+f^{(2)})_{|j|j}-(f^{(1)}+f^{(2)})_{|j}(f^{(1)}+f^{(2)})_{|j}=\nonumber\\
&&f^{(1)}_{|j|j}-f^{(1)}_{|j}f^{(1)}_{|j}+f^{(2)}_{|j|j}-f^{(2)}_{|j}f^{(2)}_{|j}
 \end{eqnarray}
Since  $f^{(1)}_{|j|j}=f^{(2)}_{|j|j}=0$ it follows that 
 \begin{equation}\label{4.6x}
f^{(1)}_{|0|0}+2f^{(1)}_{|j}f^{(2)}_{|j}=0
 \end{equation}

Take 
$$f^{(1)}=\frac{(m+iq)}{r^{(1)}}\,,\qquad f^{(2)}=\frac{(M+iQ)}{r^{(2)}}$$
 Denote $\a^{(1)}=\a, f^{(1)}=f, \qquad f^{(2)}=\phi, r^{(1)}=\rho, r^{(2)}=r.$ Thus at $y^0$ 
\begin{equation}\label{4.5}
f_{|j}=-(m+iq)(x^j-\a^j)\rho^{-3}
 \end{equation}
 \begin{equation}\label{4.6}
\phi_{|j}=-(M+iQ)x^jr^{-3}
 \end{equation}
 \begin{equation}\label{4.7}
f_{|0}=(m+iq)(x^j-\a^j)\dot{\a}^jr^{-3}
 \end{equation}
 \begin{equation}\label{4.8}
f_{|0|0}=(m+iq)(x^j-\a^j)\ddot{\a}^jr^{-3}
 \end{equation}
 where the quadratic terms in $\dot{\a}$ where omitted. At $(y^0, \a(y^0))$ the equations (\ref{4.6y}) and (\ref{4.5}---\ref{4.8}) reduce to
 \begin{equation}\label{4.9}
 \ddot{\a}^j(m+iq)f_{|j}+2(m+iq)(M+iQ)f_{|j}\phi_{|j}=0
 \end{equation}
 
 The real part of Eq. (\ref{4.9}) has to be satisfied at the vicinity of $(y^0,\a(y^0))$. The result, after cancellation of $f_{|j}$ is
  \begin{equation}\label{4.10}
m\ddot{\a}^j=-2(mM-qQ)x^jr^{-3}
 \end{equation}
 Newton and Coulomb laws, respectively. 
The approximate field equation (\ref{2.9}), for a system of two bodies, takes care of the forces.

(\ref{4.9}) is quadratic in $f_{j}$, $\phi_{j}$ and $\ddot{\a}^j$. Thus, for (\ref{4.10})  to hold, it is needed to approximate $\square_\Phi\Phi$ to the second order.   The derivation holds for $y^0$, so that $\dot{\a}(y^0)=0$, there is no consideration of the "reduced mass, e.t.c. Thus  (\ref{4.10}) is approximate. 

Let $\hat{m},\hat{M},\hat{q},\hat{Q}$ denote the masses and the charges in the M.K.S. units
  \begin{equation}\label{4.12}
\hat{m}\ddot{\a}^j=-(k\hat{m}\hat{M}-K\hat{q}\hat{Q})x^jr^{-3}
 \end{equation}
 Since $\frac d{dx^0}=\frac 1c \frac d {dt}$ it follows that $m=\frac 12 kc^{-2}\hat{m}$, $M=\frac 12 kc^{-2}\hat{M}$, $q= (kK)^{1/2}c^{-2}\hat{q}$, $Q= (kK)^{1/2}c^{-2}\hat{Q}$.
Since $k=6.67\cdot 10^{-8}$ and $K=9\cdot 10^9$ then for $M/r<10^{16}$, 
$Q/r<10^{-7}$
 \begin{equation}\label{4.14}
|1\pm exp(2(f+\phi))|=O(10^{-8})
 \end{equation}
 Thus, the approximation of the exponentials by 1 is in line with the approximations performed in this section.

 A model equation with Newton-type law of force is presented in \cite{9}.

 \section{Invariant Maxwell equations. Incorporation in a massless frame}
 For a frame $\Phi$ define $W=A^\a\Phi^\a$ to be the massless electromagnetic form. Consider the 2-form $dW$. For $\Phi^\a=dx^\a$ define 
  \begin{equation}\label{6.1}
A^0_{|j}-A^j_{|0}=E^j
 \end{equation}
 Denote by $\hat{E}$ the 2-form $E^j dx^j\wedge dx^0$. 
 Denote by $\tilde A$ the 3-dimensional vector $(A^1,A^2,A^3)$. Define $H$
  \begin{equation}\label{6.2}
curl \tilde A =H
 \end{equation}
 Denote by $\hat{H}$ the 2-form 
 $$\hat{H}=H^1 dx^2\wedge dx^3+H^2 dx^3\wedge dx^1+H^3 dx^1\wedge dx^2$$
 so that 
 \begin{equation}\label{6.2x}
dW=\hat{E}+\hat{H}
 \end{equation}
 The identity $d^2W=0$ together with the definitions (\ref{6.1}-\ref{6.2}) are equivalent to the first pair of Maxwell equations. 
  \begin{equation}\label{6.3}
curl E +\dot{H}=0,\quad div H=0
 \end{equation}
 The second pair of Maxwell equations
  \begin{equation}\label{6.4}
curl H -\dot{E}=j,\quad div E=\rho
 \end{equation}
 carries the physical content of the equations. By a straighforward computations  (\ref{6.4}) is equivalent to 
 \begin{equation}\label{6.5}
d*dW=d*(\hat{E}+\hat{H})=J
 \end{equation}
 Where the coefficients  of the 3-form $J$ are $(j,\rho)$. 
 
 For a general coordinate system define $E^j$ to be the factor of $\Phi^j\wedge\Phi^0$ in $dW$. 
 Define $H^j$ to be the factor of $\Phi^k\wedge\Phi^l$  where $(j,k,l)$ is the direct segment starting with $j$ of $(12312\cdots)$. 
The first pair of  Maxwell equations will be the identity $d^2W=0$. The second pair will be 
 \begin{equation}\label{6.8}
 d*_\Phi dW=J\,,
 \end{equation}
 where $\Phi=Re \Psi$, $\Psi$ being the complex frame that incorporates the electromagnetic field. 
 Since $*_\Phi$ is invariant so is (\ref{6.8}).  It is equivalent to Maxwell's equations. 
 
 $W$ can be incorporated into a complex linearized frame,$\Psi=\Phi+i\Lambda$. It is 
 \begin{equation}\label{6.9}
 \Psi^0_0=(1+A^0)dx^0\qquad \Psi^0_j=A^jdx^j\,,
 \end{equation}
 \begin{equation}\label{6.10}
 \Psi^j=(1-A^0)dx^j\,.
 \end{equation}
 The $A^\alpha$ are imaginary. 
 
 Indeed by (\ref{2.6}) 
 \begin{equation}\label{6.11}
 L*_\Phi\Psi^0=-\frac 13L\Psi^1\wedge\Psi^2\wedge\Psi^3=-\frac 13 (1-3A^0)dx^1\wedge dx^2\wedge dx^3\,.
 \end{equation}
  \begin{equation}\label{6.12}
  L*\Psi^0=(1+A^0)dx^1\wedge dx^2\wedge dx^3\,.
   \end{equation}
   \begin{equation}\label{6.13}
 dL*_\Phi\Psi^0=d*\Psi^0\,.
 \end{equation}
 \begin{equation}\label{6.14}
  L*_\Phi\Psi^1=L\Psi^0\wedge\Psi^2\wedge\Psi^3=(1+A^0)dx^0\wedge dx^2\wedge dx^3=L*\Psi^1\,.
   \end{equation}
   Equations (\ref{6.11}---\ref{6.14}) imply by equations (\ref{7.1}---\ref{7.5}) that 
   \begin{equation}\label{6.15}
 L\square_\Psi\Psi=\square\Psi\,.
 \end{equation}
\appendix
\section{}

The computation of $\square_\Phi\Phi$ for spherically symmetric and stationary frame. Take
\begin{equation}\label{A.1}
\Phi^0=e^{-f}dx^0\qquad \Phi^j=e^{g}dx^j
\end{equation}
\begin{equation}\label{A.2}
dx^0=e^{f}\Phi^0\qquad dx^j=e^{-g}\Phi^j
\end{equation}
The structural equations are 
\begin{equation}\label{A.3}
dH=H_{|0}dx^0+ H_{|j}dx^j=H_{|0}e^{f}\Phi^0+ H_{|j}e^{-g}\Phi^j
\end{equation}
\begin{equation}\label{A.4}
d\Phi^0=f_{|k}e^{-f}dx^0\wedge dx^k=f_{|k}e^{-g}\Phi^0\wedge \Phi^k
\end{equation}
\begin{equation}\label{A.5}
d\Phi^j=g_{|k}e^{g}dx^k\wedge dx^j=g_{|k}e^{-g}\Phi^k\wedge \Phi^j
\end{equation}

Computation of $*_\Phi d*_\Phi d\Phi^1$. 

By (\ref{A.5})
\begin{equation}\label{A.12}
*_\Phi d\Phi^1=(g_{|2}\Phi^0\wedge\Phi^3-g_{|3}\Phi^0\wedge\Phi^2)e^{-g}
\end{equation}
\begin{eqnarray}\label{A.13}
d*_\Phi d\Phi^1&=&\Big(g_{|2|1}\Phi^1\wedge\Phi^0\wedge\Phi^3+g_{|2|2}\Phi^2\wedge\Phi^0\wedge\Phi^3\nonumber\\
&&-g_{|3|1}\Phi^1\wedge\Phi^0\wedge\Phi^2-g_{|3|3}\Phi^3\wedge\Phi^0\wedge\Phi^2\nonumber\\
&&-g_{|2}g_{|1}\Phi^1\wedge\Phi^0\wedge\Phi^3-g_{|2}g_{|2}\Phi^2\wedge\Phi^0\wedge\Phi^3\nonumber\\
&&+g_{|3}g_{|1}\Phi^1\wedge\Phi^0\wedge\Phi^2+g_{|3}g_{|3}\Phi^3\wedge\Phi^0\wedge\Phi^2\nonumber\\
&&+g_{|2}f_{|1}\Phi^0\wedge\Phi^1\wedge\Phi^3+g_{|2}f_{|2}\Phi^0\wedge\Phi^2\wedge\Phi^3\nonumber\\
&&-g_{|3}f_{|1}\Phi^0\wedge\Phi^1\wedge\Phi^2-g_{|3}f_{|3}\Phi^0\wedge\Phi^3\wedge\Phi^2\nonumber\\
&&-g_{|2}g_{|1}\Phi^0\wedge\Phi^1\wedge\Phi^3-g_{|2}g_{|2}\Phi^0\wedge\Phi^2\wedge\Phi^3\nonumber\\
&&+g_{|3}g_{|1}\Phi^0\wedge\Phi^1\wedge\Phi^2+g_{|3}g_{|3}\Phi^0\wedge\Phi^3\wedge\Phi^2\Big)e^{-2g}\nonumber\\
&=&(-g_{|2|1}+g_{|2}f_{|1})e^{-2g}\Phi^0\wedge\Phi^1\wedge\Phi^3\nonumber\\
&&+[-(g_{|2|2}+g_{|3|3})+g_{|2}f_{|2}+g_{|3}f_{|3}]e^{-2g}\Phi^0\wedge\Phi^2\wedge\Phi^3\nonumber\\
&&+(g_{|3|1}-g_{|3}f_{|1})e^{-2g}\Phi^0\wedge\Phi^1\wedge\Phi^2
\end{eqnarray}
$*_\Phi d*_\Phi d\Phi^1$ read from (\ref{A.13})
\begin{eqnarray}\label{A.14}
*_\Phi d*_\Phi d\Phi^1&=&(g_{|2|1}-g_{|2}f_{|1})e^{-2g}\Phi^2+[-(g_{|2|2}+g_{|3|3})+g_{|2}f_{|2}+g_{|3}f_{|3}]e^{-2g}\Phi^1+\nonumber \\ &&(g_{|3|1}-g_{|3}f_{|1})e^{-2g}\Phi^3
\end{eqnarray}
Computation of $d*_\Phi d*_\Phi \Phi^1$. 
\begin{equation}\label{A.15x}
*_\Phi \Phi^1=\Phi^0\wedge\Phi^2\wedge\Phi^3
\end{equation}
\begin{equation}\label{A.15}
d*_\Phi \Phi^1=(f_{|1}-2g_{|1})e^{-g}\Phi^0\wedge\Phi^1\wedge\Phi^2\wedge\Phi^3
\end{equation}
\begin{equation}\label{A.16}
*_\Phi d*_\Phi \Phi^1=(f_{|1}-2g_{|1})e^{-g}
\end{equation}
\begin{equation}\label{A.17}
d*_\Phi d*_\Phi \Phi^1=\sum_{j}[(f_{|1|j}-2g_{|1|j})-(f_{|1}-2g_{|1})g_{|j}]e^{-2g}\Phi^j
\end{equation}
\begin{eqnarray}\label{A.18}
\square_\Phi \Phi^1&=&[-(2g_{|1|1}+g_{|2|2}+g_{|3|3})+f_{|1|1}-f_{|1}g_{|1}+2g_{|1}g_{|1}+f_{|2}g_{|2}+f_{|3}g_{|3}]e^{-2g}\Phi^1\nonumber\\
&&+[f_{|1|2}-g_{|1|2}+2(g_{|1}g_{|2}-f_{|1}g_{|2})]e^{-2g}\Phi^2\nonumber\\
&&+[f_{|1|3}-g_{|1|3}+2(g_{|1}g_{|3}-f_{|1}g_{|3})]e^{-2g}\Phi^3
\end{eqnarray}
The only way to  annihilate the coefficients of $\Phi^2$ and $\Phi^3$ is to take   $f=g$. Consequently
\begin{equation}\label{A.19}
\square_\Phi \Phi^1=[-g_{|j|j}+g_{|j}g_{|j}]e^{-2g}\Phi^1
\end{equation}
From now on it will be assumed that $f=g$.  

Computation of $*_\Phi d*_\Phi d\Phi^0$. 

By (\ref{A.4}) 
\begin{equation}\label{A.6}
*_\Phi d\Phi^0=(g_{|1}\Phi^2\wedge\Phi^3+g_{|2}\Phi^3\wedge\Phi^1+g_{|3}\Phi^1\wedge\Phi^2)e^{-g}
\end{equation}
\begin{eqnarray}\label{A.7}
d*_\Phi d\Phi^0&=&[(g_{|1|1}-g_{|1}g_{|1})+(g_{|2|2}-g_{|2}g_{|2})+(g_{|3|3}-g_{|3}g_{|3})]e^{-2g}\Phi^1\wedge\Phi^2\wedge\Phi^3\nonumber\\
&&+2(g_{|1}g_{|1}+g_{|2}g_{|2}+g_{|3}g_{|3})e^{-2g}\Phi^1\wedge\Phi^2\wedge\Phi^3\nonumber\\
&=&(g_{|j|j}+g_{|j}g_{|j})e^{-2g}\Phi^1\wedge\Phi^2\wedge\Phi^3
\end{eqnarray}
By the definition of $*_\Phi$ the linear terms are subject to a sign change while the quadratic terms are not. 
\begin{equation}\label{A.8}
*_\Phi d*_\Phi d\Phi^0=(g_{|j|j}-g_{|j}g_{|j})e^{-2g}\Phi^0
\end{equation}

Computation of $d*_\Phi d*_\Phi \Phi^0$.
\begin{equation}\label{A.9}
*_\Phi \Phi^0=\Phi^1\wedge\Phi^2\wedge\Phi^3
\end{equation}
\begin{equation}\label{A.10}
d*_\Phi \Phi^0=0
\end{equation}
Thus 
\begin{equation}\label{A.11}
\square_\Phi \Phi^0=(g_{|j|j}-g_{|j}g_{|j})e^{-2g}\Phi^0
\end{equation}

\section{}
The computation of $\square_\Phi\Phi$ for spherically symmetric and time dependent  frame. Take $f=g$ so that 
\begin{equation}\label{B.1}
\Phi^0=e^{-g}dx^0\qquad \Phi^j=e^{g}dx^j
\end{equation}
\begin{equation}\label{B.2}
dx^0=e^{g}\Phi^0\qquad dx^j=e^{-g}\Phi^j\,.
\end{equation}
The structural equations are 
\begin{equation}\label{B.3}
dH=H_{|0}dx^0+ H_{|j}dx^j=H_{|0}e^{g}\Phi^0+ H_{|j}e^{-g}\Phi^j
\end{equation}
\begin{equation}\label{B.4}
d\Phi^0=g_{|k}e^{-g}dx^0\wedge dx^k=g_{|k}e^{-g}\Phi^0\wedge \Phi^k
\end{equation}
\begin{equation}\label{B.5}
d\Phi^j=g_{|0}e^{g}dx^0\wedge dx^j+g_{|k}e^{g}dx^k\wedge dx^j=g_{|0}e^{g}\Phi^0\wedge \Phi^j+g_{|k}e^{-g}\Phi^k\wedge \Phi^j
\end{equation}
Computation of $*_\Phi d*_\Phi d\Phi^0$. 
\begin{equation}\label{B.6}
*_\Phi d\Phi^0=(g_{|1}\Phi^2\wedge\Phi^3+g_{|2}\Phi^3\wedge\Phi^1+g_{|3}\Phi^1\wedge\Phi^2)e^{-g}
\end{equation}
\begin{eqnarray}\label{B.7}
d*_\Phi d\Phi^0&=&[(g_{|1|0}-g_{|1}g_{|0})\Phi^0\wedge\Phi^2\wedge\Phi^3\nonumber\\
&&+(g_{|2|0}-g_{|2}g_{|0})\Phi^0\wedge\Phi^3\wedge\Phi^1\nonumber\\
&&+(g_{|3|0}-g_{|3}g_{|0})\Phi^0\wedge\Phi^1\wedge\Phi^2\nonumber\\
&&+(g_{|j|j}+g_{|j}g_{|j})e^{-2g}\Phi^1\wedge\Phi^2\wedge\Phi^3\nonumber\\
&&+2g_{|1}g_{|0}\Phi^0\wedge\Phi^2\wedge\Phi^3\nonumber\\
&&+2g_{|2}g_{|0}\Phi^0\wedge\Phi^3\wedge\Phi^1\nonumber\\
&&+2g_{|3}g_{|0}\Phi^0\wedge\Phi^1\wedge\Phi^2\nonumber\\
&=&[(g_{|1|0}+g_{|1}g_{|0})\Phi^0\wedge\Phi^2\wedge\Phi^3\nonumber\\
&&+(g_{|2|0}+g_{|2}g_{|0})\Phi^0\wedge\Phi^3\wedge\Phi^1\nonumber\\
&&+(g_{|3|0}+g_{|3}g_{|0})\Phi^0\wedge\Phi^1\wedge\Phi^2\nonumber\\
&&+(g_{|j|j}+g_{|j}g_{|j})e^{-2g}\Phi^1\wedge\Phi^2\wedge\Phi^3
\end{eqnarray}

\begin{eqnarray}\label{B.8}
*_\Phi d*_\Phi d\Phi^0&=&[(g_{|1|0}+g_{|1}g_{|0})\Phi^1+(g_{|2|0}+g_{|2}g_{|0})\Phi^2+(g_{|3|0}+g_{|3}g_{|0})\Phi^3]\nonumber\\
&&+(g_{|j|j}-g_{|j}g_{|j})e^{-2g}\Phi^0
\end{eqnarray}

Computation of $d*_\Phi d*_\Phi \Phi^0$.
\begin{equation}\label{B.9}
*_\Phi \Phi^0=-\frac 13\Phi^1\wedge\Phi^2\wedge\Phi^3
\end{equation}
\begin{equation}\label{B.10}
 d*_\Phi \Phi^0=-g_{|0}e^g\Phi^0\wedge\Phi^1\wedge\Phi^2\wedge\Phi^3
\end{equation}
\begin{equation}\label{B.11}
*_\Phi  d*_\Phi \Phi^0=-g_{|0}e^g
\end{equation}
\begin{equation}\label{B.12}
d*_\Phi d*_\Phi \Phi^0=-(g_{|0|0}+g_{|0}g_{|0}) e^{2g}\Phi^0-(g_{|0|j}+g_{|0}g_{|j}) \Phi^j
\end{equation}
Thus 
\begin{equation}\label{B.13}
\square_\Phi \Phi^0=(-g_{|0|0}e^{2g}+g_{|j|j}e^{-2g}-g_{|0}g_{|0})e^{2g}- g_{|j}g_{|j}e^{-2g}\Phi^0
\end{equation}
Computation of $d*_\Phi d*_\Phi \Phi^1$. 

\begin{equation}\label{B.17}
*_\Phi \Phi^1=\Phi^0\wedge\Phi^2\wedge\Phi^3
\end{equation}
\begin{equation}\label{B.18}
d*_\Phi \Phi^1=-g_{|1}e^{-g}\Phi^0\wedge\Phi^1\wedge\Phi^2\wedge\Phi^3
\end{equation}
\begin{equation}\label{B.18x}
d*_\Phi d*_\Phi \Phi^1=(-g_{|1|0}+g_{|1}g_{|0})\Phi^0+
(-g_{|1|j}+g_{|1}g_{|j})e^{-2g}\Phi^j
\end{equation}

Computation of $*_\Phi d*_\Phi d\Phi^1$. 

By (\ref{A.5})
\begin{equation}\label{B.14}
*_\Phi d\Phi^1=(g_{|2}\Phi^0\wedge\Phi^3-g_{|3}\Phi^0\wedge\Phi^2)e^{-g}+g_{|0}e^{g}\Phi^2\wedge\Phi^3
\end{equation}
$*_\Phi d*_\Phi d\Phi^1$ equals $*_\Phi d$ of (\ref{B.14}). 

The computation of $*_\Phi d$ of the first two terms on the right of (\ref{B.14}) is exhibited in (\ref{A.6}-\ref{A.8}).  
Together with (\ref{A.17}), (recall: $f=g$),  the sum is 
\begin{eqnarray}\label{B.17x}
[-g_{|j|j}+g_{|j}g_{|j}]e^{-2g}\Phi^1
\end{eqnarray}

 Thus 
\begin{eqnarray}\label{B.19}
*_\Phi d(g_{|0}e^{g}\Phi^2\wedge\Phi^3)=(g_{|0|0}+3g_{|0}g_{|0})e^{2g}\Phi^1+(g_{|0|1}-3g_{|0}g_{|1})\Phi^0
\end{eqnarray}

$\square_\Phi \Phi^1$ will be the sum of (\ref{B.17x}), (\ref{B.19})  and the excess of (\ref{B.18x}) over (\ref{A.17}). 
\begin{equation}\label{B.20}
\square_\Phi \Phi^1=[(g_{|0|0}+3g_{|0}g_{|0})e^{2g}+[-g_{|j|j}+g_{|j}g_{|j}]e^{-2g}]\Phi^1-2g_{|0}g_{|1}\Phi^0
\end{equation}

  
\end{document}